\documentclass{epl}
\usepackage{graphicx}%
\usepackage{amsmath}%
\usepackage{amsfonts}%
\usepackage{amssymb}
\newcommand{\be}{\begin{equation}}
\newcommand{\ee}{  \end{equation}}
\newcommand{\ba}{\begin{eqnarray}}
\newcommand{\ea}{  \end{eqnarray}}

\title{Towards a time-revesal mirror for quantum systems}
\author{H. M. Pastawski\inst{1} \and E. P. Danieli\inst{1} \and H. L. Calvo\inst{1} \and L. E. F. Foa Torres\inst{2}}

\institute{  
  \inst{1} Facultad de Matem\'{a}tica, Astronom\'{\i}a y F\'{\i}sica, Universidad Nacional de C\'{o}rdoba, Ciudad Universitaria, 5000 C\'{o}rdoba, Argentina \\
  \inst{2} CEA/DRT/LETI/DIHS/LMNO, 17 avenue des Martyrs, 38054 Grenoble, France
}

\begin{document}

\maketitle
\begin{abstract}
The reversion of the time evolution of a quantum state can be achieved by
changing the sign of the Hamiltonian as in the polarization echo experiment
in NMR. In this work we describe an alternative mechanism inspired by the
acoustic time reversal mirror. By solving the inverse time problem in a
discrete space we develop a new procedure, the perfect inverse filter. It
achieves the exact time reversion in a given region by reinjecting a
prescribed wave function at its periphery.
\end{abstract}




Performing a time reversal experiment in a quantum system seemed an
imposible task until Erwin Hahn realized that this would\ \textquotedblleft
only\textquotedblright\ require inverting the sign of the Hamiltonian. This
was achieved for a set of independent nuclear spins precessing in a magnetic
field through the sudden application of a radio frequency (rf) pulse and
became the basis of the spin echo \cite{cit--Hahn}. When a similar strategy
was applied to a system of interacting spins, a true many-body system \cite%
{cit--PolarEchoes}, the first practical realization of a Loschmidt daemon 
\cite{cit--Wheeler,cit--Kuhn} was finally achieved \cite{cit--Exper-Chaos}.
We dub this procedure a \textit{hasty daemon}, as it involves the global and
instantaneous action of a rf pulse. Notably, this action tends to be quite
dependent on the underlying instability (chaos) of the corresponding
classical system \cite{cit--JalPast}.

On the other hand, during the last decade, Fink and his group developed an
experimental technique, called time-reversal mirror (TRM), which reverse the
propagation of acoustic waves \cite{cit--FINK}. A pulse at a point source $%
\mathbf{r}_{0}$ on a working region $\mathcal{C}$ is detected as it arrives
to an array of transducers at positions $\mathbf{r}_{i}$, typically
surrounding the \textit{cavity} $\mathcal{C}$. Their registries $\psi (%
\mathbf{r}_{i},t)$ are recorded until a time $t_{R}$ at which the amplitudes 
$\psi (\mathbf{r}_{i},t)$ have become negligible. These transducers can act
alternatively as microphones or loudspeakers. Afterwards, each one re-emits
in the time reversed sequence i.e. producing an extra signal $\chi _{\mathrm{%
inj}.}(\mathbf{r}_{i},t_{R}+\delta t)=c\psi (\mathbf{r}_{i},t_{R}-\delta t)$
where $c$ is controlled by the \textquotedblleft volume\textquotedblright\
knob. The experiments show that these waves tend to refocus at the source
point at the time $t=2t_{R}$, i.e. a Loschmidt echo is formed! The
robustness of the time-reversal procedure is a prominent feature of these
experiments. Surprisingly, systems with a random scattering mechanism are
particularly stable. In fact, there is a precise prescription for time
reversal \cite{cit--Didi} that requires the control of the field function
and its normal derivative over a surrounding surface. However, the TRM
procedure is quite effective even when these requirement are not completely
satisfyied. On this basis, several applications in communications \cite%
{cit--Kuperman-OCEANcomm} and medicine \cite{cit--FINK-biolog} were already
performed. It is clear that these experiments introduce a different
procedure for time reversal: a persistent action at the periphery that we
call a \textit{stubborn daemon}. At this point the the first question to ask
is: Can this concept be applied to a Quantum Mechanical system?

A quantum experiment would touch an issue usually overlooked: \textit{the
use of the Schr\"{o}dinger equation (SE) with a time dependent source}. This
phenomenon is not merely academic since it appears in many areas: the
gradual injection of coherent polarization in the system of abundant nuclei
through\ an NMR cross-polarization transfer \cite{cit--MKBE74}; the creation
of a coherent excited state \cite{cit--Zewail} through particular sequences
of laser pulses at slow rates of pumping; and an a.c. electrical
conductivity experiment where the electrodes are fluctuating sources of
waves \cite{cit--GLBE2}. However, there is no general answer \cite%
{cit--Allcock,cit--Muga} to the \textquotedblleft inverse time
problem\textquotedblright : \textit{What wave function must be injected to
obtain a desired output? }In what follows we solve this problem for a
reasonably general case and use that solution to implement a protocol for a
perfect quantum time reversal experiment.

Let us consider a one dimensional system with a wave packet $\psi (x,t)$
localized around a point $x_{0}$ at time $t=0$ and traveling rightward. The
first question to answer is: If we record the wave function as a function of
time only at a point $\ x_{s}>x_{0}$, is it possible to use this information
to recover the same wave function? Answers to this question were given in
Ref. \cite{cit--Muga} for some particular potentials. To solve this problem
in a general way, we resort to a discrete Hamiltonian:

\begin{equation}
\widehat{\mathcal{H}}=\sum_{j}E_{j}^{{}}\widehat{c}_{j}^{+}\widehat{c}%
_{j}^{{}}+\sum_{j}(V_{x_{j+1},x_{j}}^{{}}\hat{c}_{j+1}^{+}\hat{c}%
_{j}^{{}}+h.c.),  \label{eq-Hamiltonian}
\end{equation}%
here $\hat{c}_{j}^{+}$ and $\hat{c}_{j}^{{}}$ are the creation and
annihilation operators for a particle at the coordinate $x_{j}=ja$ where $a$
is the lattice constant. In the standard notation \cite{cit--HoracioMEX} the
kinetic energy yields the hopping term $V_{x_{i},x_{j}}^{{}}=-V\delta _{i\pm
1,j}$ with $V=\hbar ^{2}/(2ma^{2})$, while the potential energy $U(x)$ fixes
the \textquotedblleft site energy\textquotedblright\ $%
E_{j}^{{}}=U(x_{j})+2V. $ We want to express $\psi (x=ja,t)=\left\langle
0\right\vert \hat{c}_{j}^{{}}\left\vert \psi (t)\right\rangle $ in terms of
the wave function at the position $x_{s}$ of the detector/source. We start
with the usual expression:

\begin{equation}
\psi(x,t)=\sum_{n}\mathrm{i}\hbar G_{x,x_{n}}^{R}(t-t_{0})\psi(x_{n},t_{0}),
\label{eq-simple-evolution-G_t}
\end{equation}
where the time retarded Green's function $G_{x,x_{n}}^{R}(t-t_{0})$
satisfies $\mathrm{i}\hbar\frac{\partial}{\partial t}G_{x,x_{n}}^{R}(t)-%
\sum_{_{i}}H_{x,x_{i}}G_{x_{i},x_{n}}^{R}(t)=\delta_{x,x_{n}}\delta\left[ t%
\right] $. In the energy representation:

\begin{equation}
\psi (x,t)=\mathrm{i}\hbar \sum_{n}\left[ \int \tfrac{\mathrm{d}\varepsilon 
}{{\small 2\pi \hbar }}\exp [\tfrac{-\mathrm{i}\varepsilon (t-t_{0})}{{\hbar 
}}]G_{x,x_{n}}^{R}(\varepsilon )\right] \psi (x_{n},t_{0}).
\label{eq-simple-evolution-G_E}
\end{equation}%
At this point, we separate the space in two portions: one will be the
working space $\mathcal{C}$\ where one intends to control the wave function.
The other, the \textit{outer region}, is the complementary infinite region
that contains the scattering states. Note that in our discrete calculation
the connection between both regions is achieved by the hopping term $%
V_{x_{s+1},x_{s}}^{{}},$ connecting the sites $x_{s}$ and $x_{s+1}$ at both
sides of the boundary. We use the Dyson equation $G_{x,x_{n}}^{R}=\bar{G}%
_{x,x_{n}}^{R}+\bar{G}%
_{x,x_{s+1}}^{R}V_{x_{s+1},x_{s}}^{{}}G_{x_{s},x_{n}}^{R}$ relating the
Green's functions $\bar{G}_{x_{j},x_{i}}^{R},$ of the semi-spaces defined by 
$V_{x_{s+1},x_{s}}^{{}}=0$, with the complete one. Hence, for$\mathrm{\,}\
x>x_{s}$ 
\begin{equation}
\psi (x,t)={\small i\hbar }\sum_{n}\int \tfrac{\mathrm{d}\varepsilon }{%
{\small 2\pi \hbar }}\exp [\tfrac{-\mathrm{i}\varepsilon (t-t_{0})}{{\hbar }}%
]\{\bar{G}_{x,x_{s+1}}^{R}(\varepsilon
)V_{x_{s+1},x_{s}}^{{}}\}[G_{x_{s},x_{n}}^{R}(\varepsilon )\psi
(x_{n},t_{0})].  \label{eq-ev-Dyson1}
\end{equation}%
The sum within square brackets can be identified with the energy
representation of the wave function ( i.e. $\psi (x_{s,}\varepsilon )$= $%
\sum_{n}{\small [}G_{x_{s},x_{n}}^{R}(\varepsilon )\psi (x_{n},t_{0})]$).
Besides, when $x>x_{s}$ the Dyson equation becomes$\
G_{x,x_{s}}^{R}(\varepsilon )$= $\left\{ \bar{G}_{x,x_{s+1}}^{R}(\varepsilon
)V_{x_{s+1},x_{s}}^{{}}\right\} G_{x_{s},x_{s}}^{R}.$ From this, we evaluate
the term in curly brackets that we replace in the Eq. (\ref{eq-ev-Dyson1})
to obtain, for $x>x_{s}$:

\begin{equation}
\psi (x,t)=\mathrm{i}\hbar \int \tfrac{\mathrm{d}\varepsilon }{{\small 2\pi
\hbar }}\exp \left[ \tfrac{-\mathrm{i}\varepsilon (t-t_{0})}{{\hbar }}\right]
G_{x,x_{s}}^{R}(\varepsilon )\frac{1}{G_{x_{s},x_{s}}^{R}(\varepsilon )}\psi
(x_{s,}\varepsilon )\,.  \label{eq--psi-semiespacio}
\end{equation}%
The SE with \ a source, $\mathrm{i}\hbar \frac{\partial }{\partial t}\psi
(x,t)-\sum_{x_{n}}\mathcal{H}_{x,x_{n}}\psi (x_{n},t)$= $\chi _{\mathrm{inj.}%
}(x,t),$ has the general solution:

\begin{equation}
\psi(x,t)=\mathrm{i}\hbar\int\tfrac{\mathrm{d}\varepsilon}{{\small 2\pi\hbar}%
}\exp\left[ \tfrac{-\mathrm{i}\varepsilon t}{{\hbar}}\right]
\sum_{x_{s}}G_{x,x_{s}}^{R}(\varepsilon)\chi_{\mathrm{inj.}%
}(x_{s},\varepsilon).  \label{eq--psi-general}
\end{equation}
This allows us to identify

\begin{equation}
\chi _{\mathrm{source}}^{{}}(x_{s},\varepsilon )=\frac{1}{%
G_{x_{s},x_{s}}^{R}(\varepsilon )}\psi (x_{s,}\varepsilon ),
\label{eq--consistent-inj}
\end{equation}%
as the Fourier transform (FT) of the function that must be injected at each
instant in order to obtain the target function. This result is valid in any
dimension and for an arbitrary potential $U(\mathbf{r})$. The condition $%
x>x_{s}$ becomes $\mathbf{r}\in $ $\mathcal{C}$, and one must interpret $%
\psi (x_{s},\varepsilon )$ as a vector whose components are the wave
amplitudes at the $N$ sites $\left\{ \mathbf{r}_{s}\right\} $ defining the
boundary $\mathcal{B}$ . Similarly, one recognizes $G_{x_{s},x_{s}}^{R}(%
\varepsilon )$ as the $N\times N$ matrix providing the correlations between
these sites. To our knowledge this is the first solution to the inverse time
problem. The key feature allowing this simple solution was the
representation of the Schr\"{o}dinger equation in a discrete basis. This
enabled a natural separation into complementary subspaces that are
re-connected through the Dyson equation.

\begin{figure}[ptb]
\includegraphics[width=10.0cm]{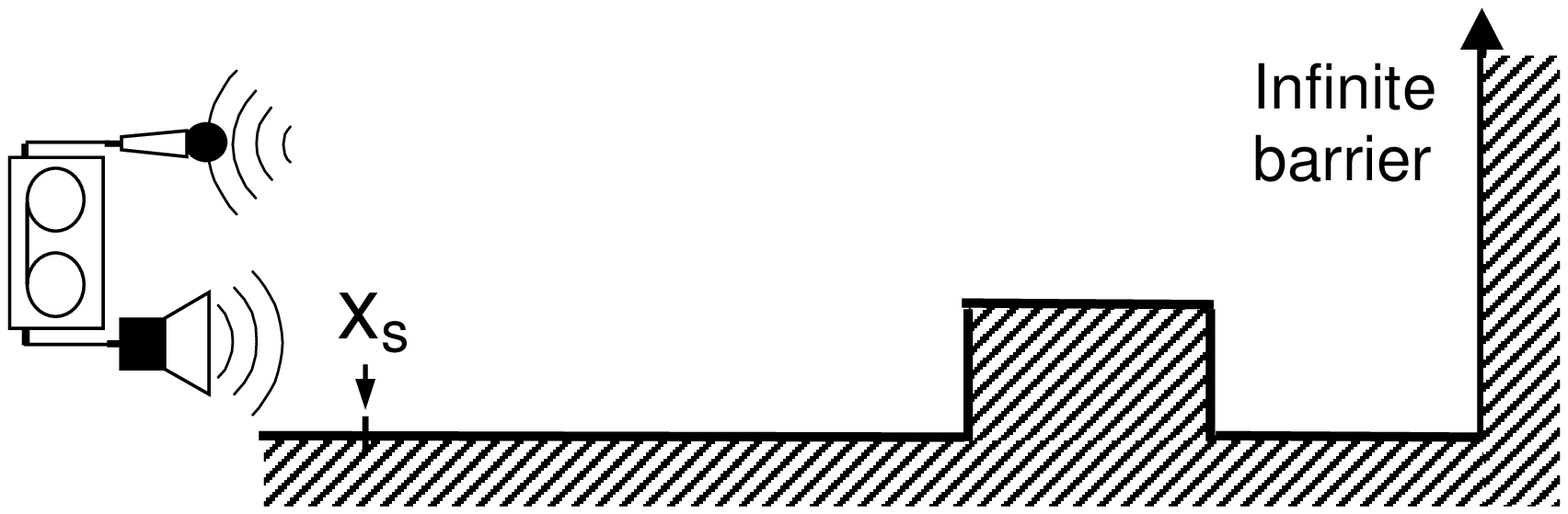} \vspace{-0.3cm}
\caption{Schematic representation of the potential profile. The barrier
height is set equal to $0.5V$ and is located between the sites $x_{s}+550a$
and $x_{s}+600a$.}
\label{fig--System}
\end{figure}

\textbf{Time-reversal via injection}. In the following, we propose a \textit{%
gedanken} scheme to achieve a perfect time-reversal of an arbitrary wave
packet by assuming that a persistent non invasive injection and detection of
waves at a \textit{single point }is possible. In such conditions one would
create an efficient stubborn daemon: the Perfect Inverse Filter (PIF). We
illustrate this by considering an incoming wave packet in a semi-infinite
space bounded by an infinite barrier at $x_{\max .}$which, together with a
scattering barrier, define a reverberant region ( see Fig. \ref{fig--System}%
). At the point $x_{s},$ located to the left of the scattering barrier, we
alternate the use of an injector and a detector of wave function
(probability and phase). This set up is a particular realization of the
\textquotedblleft sound Bazooka\textquotedblright\ scheme implemented by
Fink's group. However, instead of using the TRM, we proceed as follows:

\textbf{1)} We calculate the response of the system to an instantaneous
excitation at site $x_{s}$ i.e. $G_{x_{s},x_{s}}^{R}(t)$ and compute its FT, 
$G_{x_{s},x_{s}}^{R}(\varepsilon )$.

\begin{figure}[ptb]
\includegraphics[width=15.0cm]{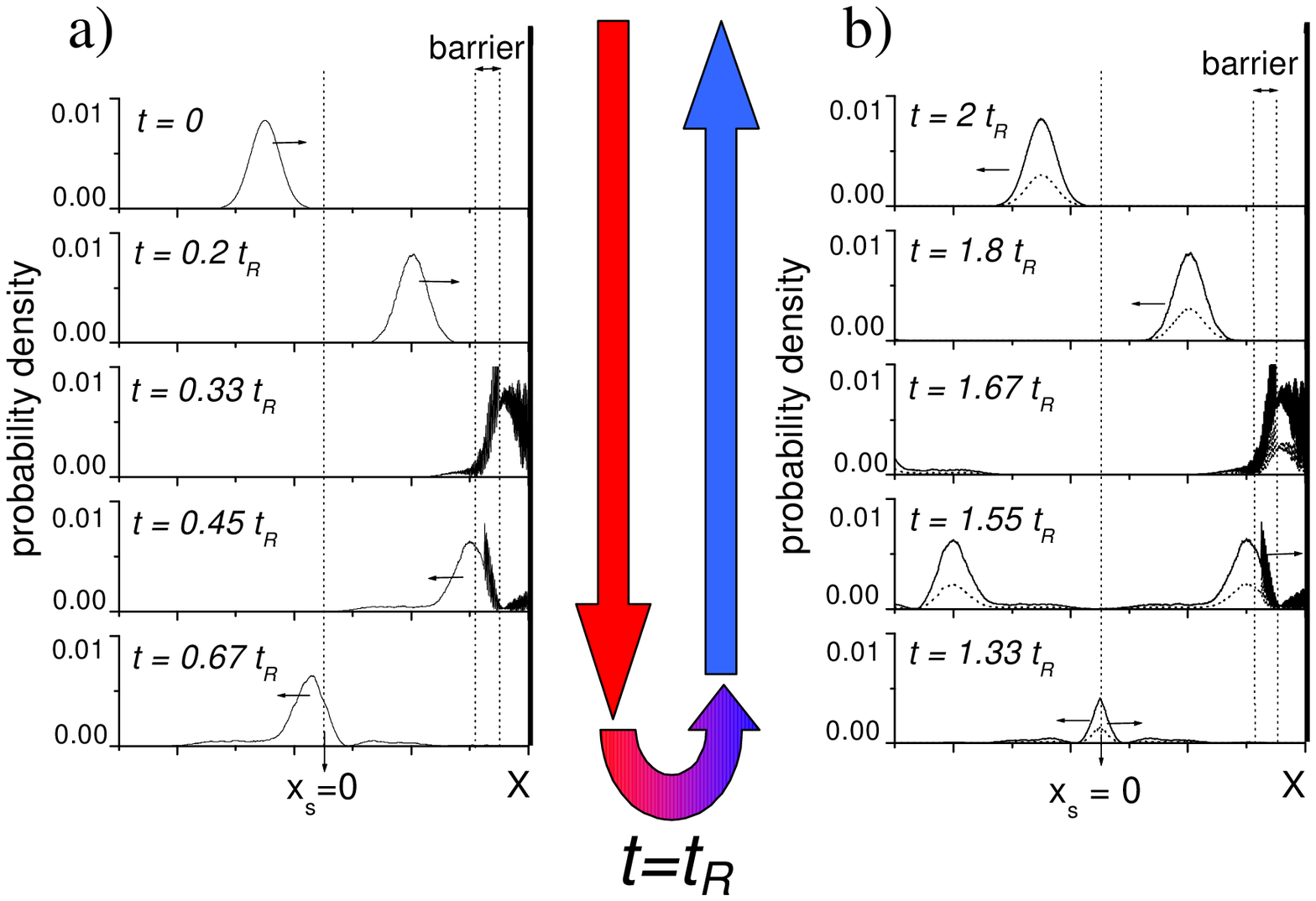} \vspace{-0.3cm}
\caption{Distribution of probability density (in the cavity and outer
region) for different times. Panel a) shows the forward evolution between $0$
and $t_{R}$. The backward evolution is shown in panel b). The solid line is
the result of an appropriate injection, Eq. (\protect\ref{eq--consistent-inj}%
), while the dotted line is obtained by injecting only the time-reversed
wave recorded at $x_{s}$ during the forward evolution. The initial gaussian
wave packet, which is centered at $x_{s}-200a$, has $\protect\sigma/a=50$
and $k_{0}a=1$. }
\label{fig--panels}
\end{figure}

\textbf{2) }We start with the empty cavity ( $\psi (x>x_{s},0)\equiv 0$) and
a wave packet that travels towards it (e.g. a Gaussian centered at $%
x_{0}<x_{s}$). The probability density at time zero is shown in the top of
the left panel of Fig. \ref{fig--panels}. It is followed by a sequence of
snapshots of the density at selected times in the range $[0,2t_{R}]$
increasing from top to bottom and continuing in the right panel from bottom
to top. The injection/detection point is indicated by a vertical dotted line
in each panel.

\textbf{3)} During the period $(0,T_{rec}^{\mathrm{PIF}}=t_{R}-t_{1})$ the
wave packet performs a free evolution: it \textit{enters} to the cavity,
collides with the barrier and then bounces back in the wall at the right end
of the system and finally \textit{escapes} towards the outer region at the
left side. See left panel of Fig. \ref{fig--panels}. Provided that there are
no localized states in the cavity and that the wave packet that escapes to
the outer region does not return, the condition $\psi (x_{s},t>t_{R})\simeq
0 $ can be fulfilled. During the whole period $T_{rec}^{\mathrm{PIF}%
}=t_{R}-t_{1}$ the wave function amplitude and phase at $x_{s}$ are \textit{%
registered}. The range $(t_{1},t_{R})$ should contain the support of $\psi
(x_{s},t),$ i.e. $\psi (x,t)\cong 0$ for $x>x_{s}$ and $t\notin
(t_{1},t_{R}).$

\textbf{4)} Now our target function is $\psi _{rev}(x_{s},t_{R}+\delta
t)=\psi ^{\ast }(x_{s},t_{R}-\delta t)$ with $0\leq \delta t\leq T_{rec}^{%
\mathrm{PIF}},$ i.e. the wave packet with reversed evolution. Using the
information registered in the previous step, we Fourier transform it

\begin{equation}
\psi _{rev}(x_{s,}\varepsilon )\simeq \int_{0}^{T_{rec}^{\mathrm{PIF}}}\psi
^{\ast }(x_{s},t_{R}-\delta t)\exp [\tfrac{\mathrm{i}\varepsilon
(t_{R}+\delta t)}{{\hbar }}]\mathrm{d}\delta t.  \label{eq--psi-rev-E}
\end{equation}%
and normalize it according to Eq. (\ref{eq--consistent-inj}). Transforming
back to time we get the actual time dependent injection acting for a time $%
T_{rec}^{\mathrm{PIF}}$. The injection also produces a wave packet that
travels to the left, i.e. escaping to the left outer region, see Fig. \ref%
{fig--panels}. Hence, perfect time reversion is restricted to the cavity,
i.e. $x>x_{s}$.

\textbf{5)} After injection has seased, the \textit{original}\textbf{\ }%
wave-packet is recovered at time $2t_{R}$ with an inverted momentum: this is
the Loschmidt Echo. Figure \ref{fig--panels} also shows, in dotted line, the
echo resulting from TRM procedure \cite{cit--FINK}, which in this case would
require the recording only the outgoing wave described in step \textbf{3)}
which is time-reverted and reinjected without further procesing.

\begin{figure}[tbp]
\includegraphics[width=12.0cm]{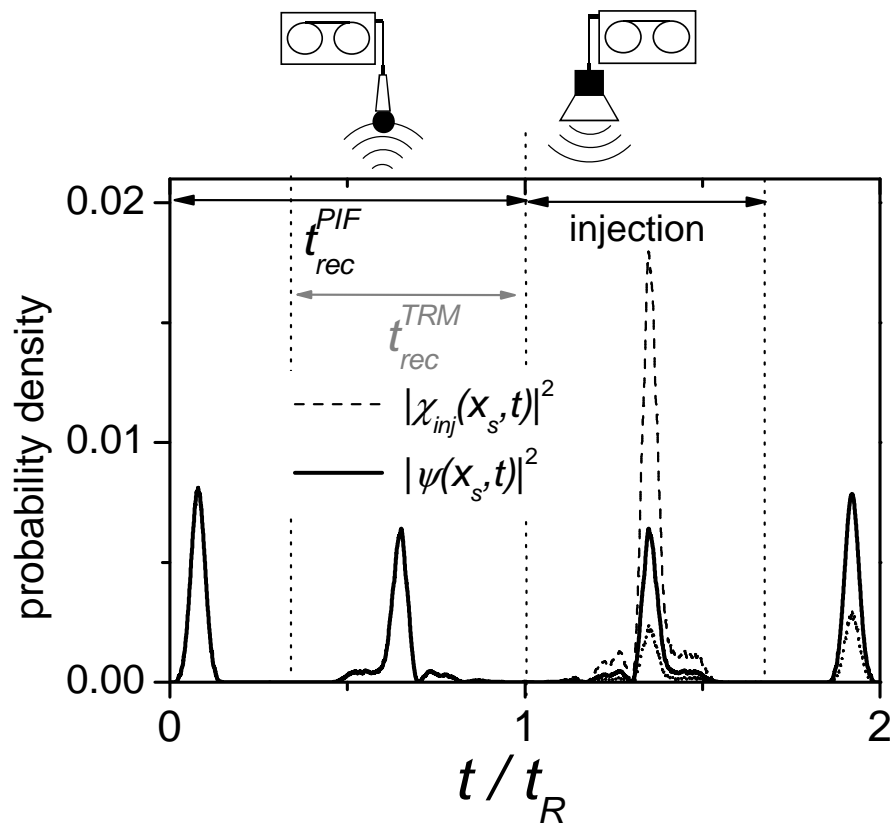} \vspace{-0.3cm}
\caption{Probability density at site $x_{s}$ as a function of time. The
actual density (solid line) results from the injection of the density shown
with a dashed line corresponding with the FT of Eq.(\protect\ref%
{eq--consistent-inj}). The density corresponding to the injection of the
time-reversal of the recorded amplitude is also shown with a dotted line.
The injection and recording periods, at site $x_{s}$, are specified in the
top of the figure.}
\label{fig--Injection}
\end{figure}

\begin{figure}[tbp]
\includegraphics[width=12.0cm]{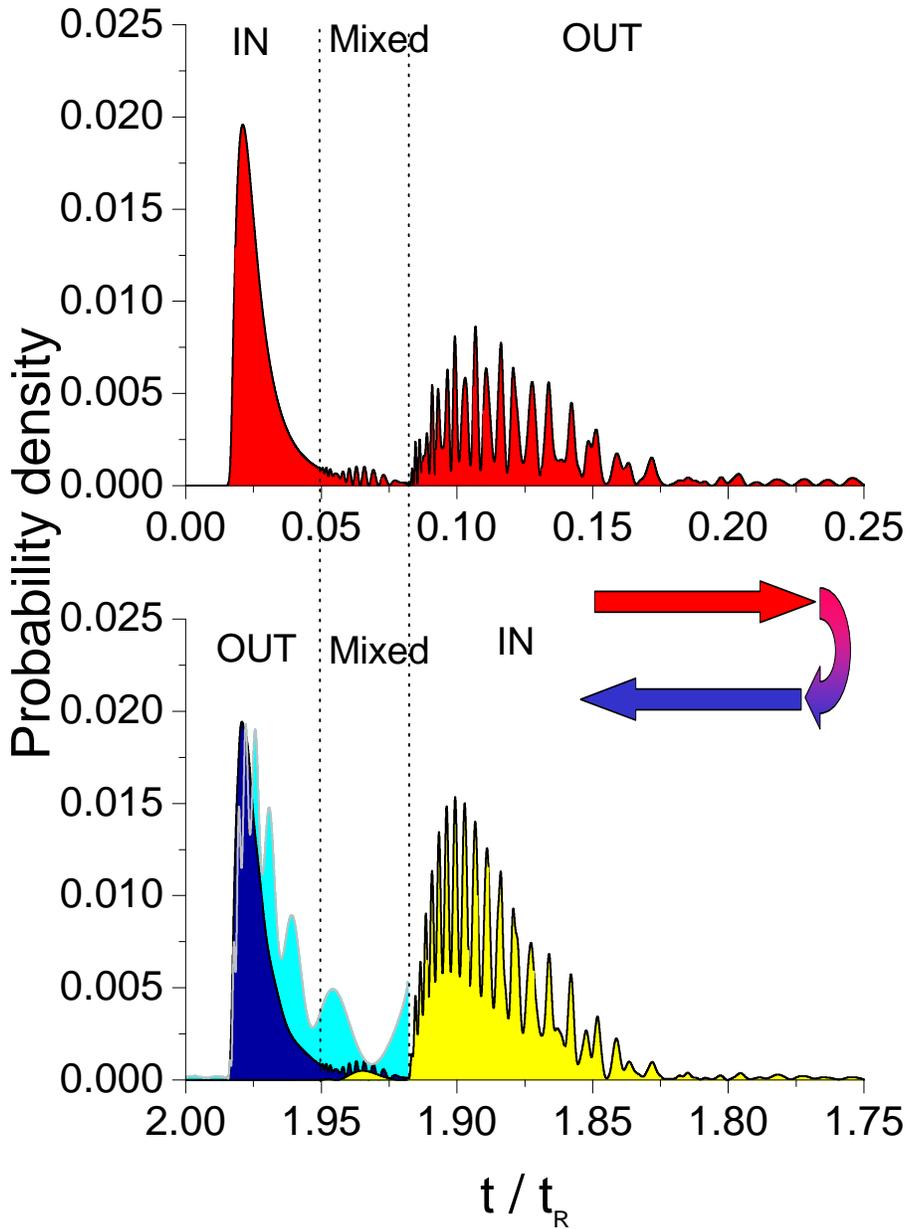} \vspace{-0.3cm}
\caption{Upper panel: density at $x_{s}$ in a forward evolution of a wave
packet. Lower panel: At rigth, the density injected at $x_{s}$ by the PIF
procedure (shaded light-grey/yellow on-line) . Its evolution determines the
correct reversed density (shaded dark/blue on line). The density obtained
from TRM is shaded grey/cyan on-line. System parameters: Barrier height $0.2V
$ between sites $x_{s}+100a$ and $x_{s}+105a$ and $x_{\max .}=200a.$
Gaussian wave packet centered at $x_{s}-100a$, with $\protect\sigma /a=3$
and $k_{0}a=0.8$. }
\label{fig4}
\end{figure}

In Fig. \ref{fig--Injection}, the density at site $x_{s}$ is shown at
different times. The actual PIF density is plotted with a solid line while
the injected density at each time is shown with a dashed line. Notice that
the injection intensity has to provide a wave propagating toward both the
cavity and outer region. We also show, with a dotted line, the density
obtained from the TRM procedure. Note that this density also exhibits an
Echo at time $2t_{R}$ but with a reduced amplitude as compared with the
original signal. While the PIF and TRM densities differ in their magnitude,
their shape in $[t_{R},2t_{R}]$ is remarkably simmilar. This indicates the
stability of the TRM in the condition considered. We emphasize that by
injecting probability amplitude according to Eq.(\ref{eq--consistent-inj})
we \textit{exactly} reverse the forward evolution of the initial wave
packet. The correction imposed by the PIF procedure becomes non-trivial in
cases where the incoming and outgoing signals superpose. In these cases the
PIF procedure \textquotedblleft filters\textquotedblright\ the outgoing
portion as can be appreciated in Fig. \ref{fig4}. In the upper panel we
display the forward evolution as registered at site $x_{s}$. In this case,
one can roughly indentify three time regimes: entrance (IN), escape (OUT)
and a Mixed region when both components interfere. The lower panel displays
the time reversal procedure. The TRM would inject only the OUT portion of
the registry shown in the upper panel. In contrast, the PIF procedure yields
an injection that extends to the mixed region shown ligth grey shaded
(yellow area on-line). The dark shaded (blue area on-line) PIF intensity
constitutes a substantial improvement over the gray shaded (cyan area
on-line) TRM signal.

The PIF protocol is valid for the reversal of any scalar waves as long as
they satisfy a linear equation. Then, different propagators are described by
the Green's functions. The basic ingredients apply to elastic or
electromagnetic waves \cite{cit--Baranger} extending the range of
applicability of the concepts introduced here.

The implementation of a stubborn Loschmidt daemon in a quantum system is not a
simple task. However, standard pulsed NMR has the tools. In an ensemble of
linear molecules, the interactions between nuclear spins can be manipulated
\cite{cit--Madi} to obtain a polarization which is the square modulus of a
single particle wave function or polarization amplitude \cite{PRL95} and
constitutes a pseudo-pure state \cite{cit--XY-dynamics}. Detection at each
time involves an ensemble measurement and a new experiment. In order to
generate a local source/detector one resorts to the interaction between
different nuclei, which can be engineered at will, e.g. a $^{13}$C \ acts as
such probe for a labeled $^{1}$H. In fact, we have been able to inject a wave
packet in a $^{1}$H ring and follow its dynamics detecting simultanously the
amplitude and relative phase at the labeled $^{1}$H \cite{JChPhys98}. This is
a double-slit like experiment that allows interferometry in the time domain
\cite{ChemPhysLett96}. If most of the polarization stays in the $^{13}$C, it
is the small portion transfered to the proton system the one described by the
theory above. While a full implementation of the quantum TRM or PIF requires
setting many important experimental details, every step towards that goal
would have potential use in spectral edition and quantum information
processing. More immediately, classical wave systems, could benefit from our
procedures which can be incorporated in a straightforward manner.

In summary, we have studied the Schr\"{o}dinger equation with source boundary
conditions. We have obtained a general solution for the inverse time problem
which is expressed in terms of the Green's function at the boundary region.
Our results enabled us to develop the Perfect Inverse Filter protocol to
implement\ a stubborn Loschmidt daemon. It allows one to achieve the perfect
time reversal of the wave dynamics obtaining a Loschmidt Echo. In some cases,
this protocol could improve the experimental procedure implemented with sound
waves. Now that a perfect reversion can be obtained, a number of questions
relevant to the Quantum Chaos field become pertinent related to the assestment
of infidelity sources\textit{.} On the view of the experimental results in
sound waves \cite{cit--FINK}, a stubborn daemon yields more robust results
than its hasty counterpart. Hence, a whole field of study opens up.

\newpage

\end{document}